\documentclass[english,floatfix,showkeys,superscriptaddress,nofootinbib,aps,prd]{revtex4}
\usepackage[latin1]{inputenc}
\usepackage[T1]{fontenc}
\usepackage{lmodern}
\usepackage{amsmath}
\usepackage{amssymb}
\usepackage{graphicx}
\usepackage{esint}
\usepackage{longtable}
\usepackage{dcolumn}
\usepackage{babel} 
\usepackage{csquotes} 
\usepackage{color} 

\begin{document}

\title{Stars and quark stars in bumblebee gravity}

\author{Juliano C. S. Neves}
\email{juliano.neves@unifal-mg.edu.br}
\affiliation{Instituto de Ciência e Tecnologia, Universidade Federal de Alfenas, \\ Rodovia José Aurélio Vilela,
11999, CEP 37715-400 Poços de Caldas, MG, Brazil}

\author{Fernando G. Gardim}
\email{fernando.gardim@unifal-mg.edu.br}
\affiliation{Instituto de Ciência e Tecnologia, Universidade Federal de Alfenas, \\ Rodovia José Aurélio Vilela,
11999, CEP 37715-400 Poços de Caldas, MG, Brazil}

\begin{abstract}
In this work, the interior spacetime of stars is built in a Lorentz symmetry breaking model called bumblebee gravity. Firstly, we calculated the modified Tolman-Oppenheimer-Volkoff equation in this context of modified gravity. Then we show that the bumblebee field, responsible for the symmetry breaking, increases the star mass-radius relation when it assumes its vacuum expectation value. When compared to the general relativity mass-radius relation, a Lorentz symmetry breaking context, like the bumblebee gravity, could provide more massive stars, surpassing the $2.5 M_{\odot}$ limit as the interior of the star is described by quark matter with the MIT bag model. 
Also, we investigate the stability of the solution with the MIT bag equation of state in this 
context of modified gravity.
\end{abstract}


\maketitle

\section{Introduction}

The violation of Lorentz symmetry has attracted increasing attention from researchers in 
recent years \cite{Liberati2013, Berti:2015itd, Addazi:2021xuf,Berti:2018cxi}. 
The central idea is that the effects of this violation, 
which concerns a fundamental symmetry in physics, manifest in high-energy contexts 
such as string theory \cite{Kostelecky1989,Kostelecky1989a,Kostelecky1989b,Kostelecky1991}, 
warped brane worlds \cite{Santos:2012if}, noncommutative geometries \cite{Carroll:2001ws}, 
and Ho$\breve{{\rm r}}$ava-Lifshitz gravity \cite{Horava2009}. In particle physics, 
the Standard Model extension (SME), developed 
by Colladay and Kostelecký \cite{Kostelecky1997,Kostelecky1998}, is conceived of as an effective 
framework in which Lorentz symmetry is broken. Various sectors of the SME have been extensively studied, 
focusing primarily on placing more stringent upper bounds on the parameters that measure 
Lorentz symmetry violation. Within SME approaches, CPT symmetry violation is explored, 
for example, in Refs. \cite{Bluhm:1997qb,Bluhm:1997qbc,Kostelecky:1999mr,Bluhm:1999dx}, 
the fermion sector in Refs. \cite{Kostelecky:2000mm,Colladay:2001wk,Altschul:2004xp}, 
the CPT-odd/even gauge sectors in Refs. \cite{Carroll:1989vb,Andrianov:1998wj,Araujo:2023izx}, 
and radiative corrections in Refs. \cite{Jackiw:1999yp,Perez-Victoria:1999erb,Altschul:2004gs,Nascimento:2007rb,Belchior:2023cbl}.
  
However, when gravity is at stake, one should take into account a mechanism of spontaneous 
symmetry breaking  \cite{Bluhm2005,Bluhm2008}  in order to maintain the geometric constraints 
and conservation laws  from general relativity. In this sense, when gravity is not weak, 
bumblebee models provide the Lorentz symmetry breaking (spontaneously  broken) as
 the bumblebee field (a vector or a tensor field) reaches its nonzero vacuum expectation value (VEV). 
In the gravitational sector, bumblebee models are studied in 
curved spacetimes \cite{Kostelecky2004, Bailey2006,Filho:2022yrk},
 gravitational waves \cite{Maluf:2014dpa,Kostelecky:2016kkn,Kostelecky:2016kfm,Amarilo:2023wpn}, black hole physics \cite{Bertolami,Casana,Euclides2020,Maluf_Neves},
wormholes \cite{Ovgun2020,Rondineli2019}, and cosmology \cite{Capelo:2015ipa,ONeal,Petrov,Neves:2022qyb,Maluf:2021eyu,Maluf:2021lwh}.  

In this work, we computed the interior geometry of a star or quark star in the bumblebee context.  
Quark stars are important because, in conditions of extreme energy density, the strong nuclear force results 
in matter composed of quarks and gluons, as evidenced by relativistic 
heavy-ion collisions~\cite{Gardim:2019xjs}. Recent data suggests that observations of compact stars 
correspond with the description of quark stars, or at least, are consistent with quark cores 
within neutron stars \cite{Annala:2019puf}. For that purpose, 
we assume that the Casana \textit{et al.} \cite{Casana} solution, also called Schwarzschild-like black hole, 
describes the exterior of a static and noncharged star in bumblebee gravity. Like the general relativity case, in which the 
interior solution matches the Schwarzschild metric at the radius of the star, in the case studied here 
we adopt the same strategy to build the interior solution. The modified Tolman-Oppenheimer-Volkoff  (TOV) 
equation is calculated. Whether in general relativity or in the bumblebee model, the TOV equation 
provides the  hydrostatic equilibrium description of fluids inside compact objects like neutron stars or quark stars. 
Two types of fluid will be used to describe the matter inside the star and to solve the modified TOV. 
In the first scenario, we employ a fluid with a constant density (incompressible) as a didactic case. 
The second one is the quark matter described by the MIT bag model, 
something more \textit{realistic} than the first case. 

The MIT bag model \cite{Chodos:1974je} is a simplified way to describe how quarks and gluons are confined 
in protons and neutrons. It is based on the idea that the energy of this state comes from the lowest energy 
state of the quarks and gluons, plus the vacuum of this state, which is different from the vacuum outside a small region. 
We refer to this energy as the bag constant,  $\mathcal{B}$, representing the energy within the bag (a small region) that 
confines quarks and gluons. In turn, this same energy generates a confinement force, balancing the 
pressure of the quarks inside the bag with the pressure of the external vacuum. With this description, 
one can construct an equation of state in which there is a contribution of this vacuum energy. 
Although this model is a phenomenological description and does not account for the 
details of quantum chromodynamics interaction, it can aid in understanding qualitatively the 
behavior of matter under extreme density conditions, such as in compact stars like neutron stars and 
quark stars \cite{Itoh:1970uw, Farhi:1984qu, Witten:1984rs, Fraga:2013qra}.

In this sense, quark matter in astrophysics has been subject of many papers. In modified theories of gravity, we see works in Einstein-Gauss-Bonnet gravity \cite{Tangphati:2021tcy,Tangphati:2021mvu,Pretel:2021czp,Gammon:2023uss,Pretel:2021gqq}, 
$f(R,L_{m},T)$ gravity \cite{Tangphati:2024war}, Weyl geometric gravity \cite{Haghani:2023nrm} 
and many others (see Ref. \cite{Olmo:2019flu} for a review). 
Quark stars are studied in modified gravity due to the possibility of having larger mass-radius relations 
in these theories  (when compared to general relativity). 
Recent data of gravitational waves events, like GW 190814 \cite{LIGO}, 
points out that one of the components of the coalescence between a black hole and a compact 
star presents mass $M=2.59^{+0.08}_{-0.09} M_{\odot}$, where $M_{\odot}$ is the solar mass. 
However, using the MIT bag model and reasonable values for the bag constant (as we will see), 
general relativity is not able to surpass the $2.4 M_{\odot}$ value for a static and noncharged star. 
On the other hand, a star in the bumblebee gravity could have $M=2.53 M_{\odot}$ or even larger values.
It is worth pointing out that the nature of the compact object in the GW190814 event is still
unknown. Besides the quark star hypothesis \cite{Tangphati:2024war,Haghani:2023nrm},
 there are alternatives that describe such a 
compact object as a light black hole \cite{LIGO,Nathanail:2021tay}, a fast rotating neutron star  \cite{Zhang:2020zsc,Most:2020bba,Biswas:2020xna}
or even a hybrid star \cite{Zhang:2020dfi}. Another alternative is 
to describe the compact object in the GW190814 event as 
a neutron star in the modified context of the Eddington-inspired Born-Infeld theory  \cite{Prasetyo:2021kfx}.
In such a context, a neutron star with $2.6 M_{\odot}$ is possible.
However, as mentioned, we assume here that the compact object in the GW190814 event could be a
quark star in bumblebee gravity.

Recently, three works on bumblebee gravity and stars drawn our attention.
The first one \cite{Ji:2024aeg} builds an interior solution from another exterior solution in bumblebee gravity, namely the spacetime obtained in Bertolami and Páramos \cite{Bertolami}. 
In our case, as mentioned before, the exterior solution is the Schwarzschild-like metric of Casana \textit{et al.} \cite{Casana}. In this case, due to the fact that different exterior solutions generate different interior solutions,
 there are no overlapping results. Regarding the two other papers, the authors tried to obtain the interior solution of a star from the Schwarzschild-like metric like us. But the modified TOV is incorrect in either cases. 
 In Ref. \cite{Liu}, for example, the polytropic fluid studied is not  solution of all modified field equations, and in Ref. \cite{Panotopoulos:2024jtn} even the exterior solution (a vacuum solution) 
  is not solution of the field equations obtained in the mentioned work.  

This article is structured as follows: In Section II the adopted bumblebee model is presented, 
the exterior solution, Schwarzschild-like spacetime, is briefly discussed. 
The analytic interior solution is built. Section III discusses two types of fluid for solving the modified TOV equation. 
In the incompressible case, approximate expressions for the pressure and the metric inside the star are shown. 
The second type of fluid under study is the quark matter that the MIT bag model describes. 
The mass-radius result is exhibited, and we can see the enhancement of this relation due to the bumblebee field. 
In Section IV, stability of the solution with the MIT bag equation of state is studied calculating
the speed of sound in the fluid, the adiabatic index and the mass-central pressure diagram.
The final comments are in Section V. In our calculations we adopt geometrized units, i.e., $G=c=1$, where $G$ 
is the Newtonian constant, and $c$ is speed of light in vacuum. As we present the results, 
the units in terms of MeV/fm$^3$ are restored.

\section{Stars in bumblebee gravity}

\subsection{The adopted model}

Bumblebee models are \textit{simple} mechanisms for studying, in the gravitational sector, the spontaneous 
breaking of (arguably) the most important symmetry in physics, the Lorentz symmetry. 
As mentioned in Introduction, these models have a nontrivial 
VEV, which could affect the dynamics of other fields coupled to the bumblebee field. Also 
bumblebee models preserve both geometric structures and conservation laws compatible with  
 general relativity \cite{Kostelecky2004,Bluhm2005}. 
 
The model adopted here regards a vector field $B_{\mu}$, the bumblebee field (in a torsion-free spacetime),
given by the following action:
\begin{equation}
 S_{B}=\int d^{4}x\sqrt{-g} \bigg [ \frac{R}{2\kappa}+\frac{\xi}{2\kappa}B^{\mu}B^{\nu}R_{\mu\nu}-\frac{1}{4} B_{\mu\nu}B^{\mu\nu} -V(B^{\mu}B_{\mu}\pm b^{2})+\mathcal{L}_{M} \bigg ],
 \label{S1} 
\end{equation}
 where $\kappa=8\pi G/c^4$ is the gravitational coupling constant, 
 and $\xi$ (with mass 
 dimension $[\xi]=M^{-1}$) plays the role of a coupling constant that accounts for the nonminimum interaction 
 between the bumblebee field and the geometry \cite{Bailey2006}. Another important
 ingredient in the action is the field strength, defined as
\begin{equation}
B_{\mu\nu}\equiv\partial_{\mu}B_\nu-\partial_{\nu}B_{\mu}.
\end{equation}
 The Lagrangian  $\mathcal{L}_{M}$ describes the matter content, which will be treated as a perfect fluid in the 
 next section. 
 Lastly, the potential $V$ is responsible for triggering spontaneously the Lorentz violation as the bumblebee 
 field assumes a nonzero
  VEV $\left\langle B_{\mu}\right\rangle \equiv b_{\mu}$, 
obeying the condition $B^{\mu}B_{\mu}=\mp b^{2}$. It is worth emphasizing that the constant $b^{2}$ 
is a real (thus positive) number, and the $\pm$ sign in Eq. (\ref{S1}) 
means that $b_{\mu}$ is timelike or spacelike, respectively, in 
agreement with the metric signature adopted in this article, namely  $(-,+,+,+)$.

The modified gravitational field equations in the bumblebee context are directly obtained  by varying the 
action (\ref{S1}) with respect to
the metric tensor $g_{\mu\nu}$ (as the bumblebee field $B_{\mu}$ is held fixed). 
This calculation leads to 
\begin{align}
G_{\mu\nu}  = & \ R_{\mu\nu}-\frac{1}{2}g_{\mu\nu}R = \kappa \left( T^{B}_{\mu\nu}+ T^{M}_{\mu\nu} \right) \nonumber \\
  = & \ \kappa\left[2V'B_{\mu}B_{\nu} +B_{\mu}^{\ \alpha}B_{\nu\alpha}-\left(V+ \frac{1}{4}B_{\alpha\beta}B^{\alpha\beta}\right)g_{\mu\nu} \right]  +  \xi \bigg[ \frac{1}{2}B^{\alpha}B^{\beta}R_{\alpha\beta}g_{\mu\nu}-B_{\mu}B^{\alpha}R_{\alpha\nu}  -B_{\nu}B^{\alpha}R_{\alpha\mu}  \nonumber\\
  &  + \frac{1}{2}\nabla_{\alpha}\nabla_{\mu}\left(B^{\alpha}B_{\nu}\right)+\frac{1}{2}\nabla_{\alpha}\nabla_{\nu}\left(B^{\alpha}B_{\mu}\right)   -  \frac{1}{2}\nabla^{2}\left(B_{\mu}B_{\nu}\right)-\frac{1}{2}
g_{\mu\nu}\nabla_{\alpha}\nabla_{\beta}\left(B^{\alpha}B^{\beta}\right) \bigg]  +\kappa T^{M}_{\mu\nu}, 
\label{modified}
\end{align}
The tensor $G_{\mu\nu}$ is the usual Einstein tensor, and the operator $'$ means derivative with respect 
to the potential argument (consequently, throughout this article $'$ is also derivative 
with respect to the radial coordinate), and $\nabla_\mu$ is the covariant derivative.
 The bumblebee energy-momentum tensor and the matter energy-momentum tensor are
$T^{B}_{\mu\nu}$ and $T^{M}_{\mu\nu}$,  respectively. 
In order to solve Eq. (\ref{modified}) and to build a spacetime metric, 
we should  choose a bumblebee potential $V$
 and a metric \textit{Ansatz} with the desired spacetime symmetry. Here, we study static stars, therefore
 we should adopt  the the general spherically symmetric \textit{Ansatz}.
 From these choices,  a set of equations is obtained, and a full metric is written. 
 In our case, the interior metric of a star. 
 
 The bumblebee model action (\ref{S1}) also gives us the equation of motion for  the 
 field $B_{\mu}$. That is, by varying (\ref{S1}) with
 respect to the field $B_{\mu}$, one has
 \begin{equation}
 \nabla_{\mu}B^{\mu\nu}= J_{B}^{\nu}+J_{M}^{\nu},
 \label{B_eq}
 \end{equation}
where $J_{M}^{\nu}$, due to the matter content,  is some sort of source for the bumblebee field, and 
\begin{equation}
J_{B}^{\nu} = 2\left( V'B^{\nu}-\frac{\xi}{2\kappa}B_\mu R^{\mu\nu}  \right)
\end{equation}
 is the term that describes the self-interaction of the field $B_{\mu}$.

\subsection{The exterior solution}

In Casana \textit{et al.} \cite{Casana}, a vacuum and static solution in the bumblebee context was built 
from the spherically symmetric \textit{Ansatz}
in the $(t,r,\theta,\phi)$ coordinates, i.e., 
\begin{equation}
g_{\mu\nu}=\mbox{diag}(-e^{2\alpha(r)}, e^{2\beta(r)},r^2, r^2\sin^2\theta).
\label{Ansatz}
\end{equation}
The authors also have chosen the following radial 
bumblebee field
\begin{equation}
B_\mu=b_{\mu}=(0,b_{r}(r),0,0),
\label{VEV}
\end{equation}
where
\begin{equation}
b_{r}(r) =\left\vert b\right\vert e^{\beta(r)}.
\label{bb}
\end{equation}
From those choices, we see that the condition $b_{\mu}b^{\mu}=b^2=\mbox{constant}$ is fully
satisfied. It is worth noting that the vacuum condition of the field
 (\ref{VEV}) leads then to a bumblebee field frozen in its VEV, that is, 
$V=0$, and the assumption $V'=0$ guarantees  that the field keeps on the minimum of the potential.
Another consequence of that particular field choice is that the field strength is zero ($B_{\mu\nu}=0$). 

From both the \textit{Ansatz}  (\ref{Ansatz}) and the field (\ref{VEV}), 
according to Ref. \cite{Casana}, 
the metric  able to describe a static and neutral black hole in vacuum (or alternatively the exterior region of a nonrotating star) in the bumblebee context is
\begin{equation}
ds^{2}=-\left(1-\frac{2M}{r}\right)dt^{2}+(1+\ell) \left(1-\frac{2M}{r}\right)^{-1}dr^{2} +r^2 \left(d\theta^2+ \sin^2\theta d\phi^2 \right).
  \label{Casana}
\end{equation}
That is to say, the spacetime described by Eq. (\ref{Casana}), also called Schwarzschild-like spacetime, is a
spherically symmetric solution of the modified field 
equations (\ref{modified}) in vacuum, that is, $T_{\mu\nu}^M=0$. The parameter $M$ is conceived of
as the black hole mass (or stellar mass in this article).
The Lorentz symmetry is broken by the \textit{Lorentz-violating parameter}
$\ell=\xi b^2$, it is the presence of the bumblebee field in the geometry. 
Of course, for $\ell=0$, the Schwarzschild metric is restored in the general relativity context.
Also, Casana \textit{et al.}, in the 
same work, evaluated the light deflection, time delay of light, and the 
perihelion advance of the planet Mercury in order to obtain
upper bounds on the parameter $\ell$. Accordingly, the most stringent upper bound is
$\ell < 10^{-15}$. However, such a limit comes from Solar system experiments.
In this article, due to high pressures and densities, we are going to relax this upper bound, allowing
 then higher values for the Lorentz-violating parameter.
 
Lastly, as pointed out in \cite{Casana}, the bumblebee field acts on the geometry modifying measurable parameters. However, as the event horizon localization is given by $g^{rr}=e^{-2\beta}=0$, we can see that the event horizon  radius $r_{+}=2M$ is the
 same in the Schwarzschild and Schwarzschild-like spacetimes.

\subsection{The interior solution}

Contrary to the exterior solution, the interior spacetime  is not described by a vacuum solution.
Here, for the sake of simplicity, we adopt the matter content inside a star given by the perfect fluid tensor
\begin{equation}
T_{\mu\nu}^{M}= \left(\rho + p \right)u_{\mu}u_{\mu} + pg_{\mu\nu},
\label{T}
\end{equation}
where $\rho$ and $p$ are the energy density and pressure of the fluid, respectively,
they are $r$-dependent functions. Also, the vector
$u_{\mu}$ is the four-velocity of the fluid, which satisfies the normalization $u_{\mu}u^{\mu}=-1$. 
Above all, the interior solution described by a perfect fluid should match the vacuum 
exterior solution as $r=R$, where $R$ is the radius of the star. 
Due to the field (\ref{VEV}), $B_{\mu\nu}=0$, and due to the matter content, 
one has  $J_{B}^{\mu}=-J_{M}^{\mu}$ from Eq. (\ref{B_eq}). 
However, for the two cases studied here, we calculate that 
$\vert J_{B}^{\mu} \vert \approx 0$, consequently we assume that
the matter sector is not coupled with the bumblebee field. Thus, the perfect fluid description (\ref{T})
is still viable.

With the metric \textit{Ansatz} (\ref{Ansatz}) and the matter description (\ref{T}), 
the modified field equations (\ref{modified}) are written as 
\begin{align}
 \frac{e^{-2\beta}}{r^2}\bigg[& e^{2\beta}-(1+\ell)\left(1-2r \beta' \right)  \bigg] = \kappa \rho, \label{Eq00} \\
  \frac{e^{-2\beta}}{r^2} \bigg[& (1+\ell) \left(1+2r\alpha' \right)- e^{2\beta} - \ell r^2 \Big(\alpha'' +\alpha'^2  
 -\alpha' \beta' - \frac{2}{r} \beta' \Big)  \bigg]  = \kappa p, \label{Eq11} \\
  (1+\ell)e^{-2\beta} \bigg[& \alpha''+\alpha'^2-\alpha'\beta'+\frac{1}{r}\left(\alpha'-\beta' \right)\bigg] = \kappa p.\label{Eq22} 
\end{align}
It is worth noticing that, due to $G_{\phi\phi}=\sin^2\theta G_{\theta\theta}$, one has just three independent equations from Eq. (\ref{modified}). Another equation comes from the local energy conservation equation $\nabla_{\nu}T^{\mu\nu}=0$. The 
component $\mu=r$ is not identically zero and will provide the modified TOV. First all, following the general relativity case, we assume that the metric component $g_{rr}$ for  the exterior solution is
\begin{equation}
g_{rr}= e^{2\beta}=(1+\ell)\left(1-\frac{2m(r)}{r} \right)^{-1},
\label{grr}
\end{equation}
and the mass function in the spherically symmetric case is
\begin{equation}
\frac{dm}{dr}=4\pi r^2 \rho.
\label{mass}
\end{equation}
In order to match the exterior solution, one assumes both $m(R)=M$ and $m(0)=0$. 
With the metric component (\ref{grr}), and 
with the aid of  Eq. (\ref{Eq22}), the mentioned component of the energy-momentum equation conservation yields  
\begin{equation}
\frac{dp}{dr}=-\left(\frac{\rho + p +\ell \left(\rho+2p \right)}{1+2\ell} \right)\alpha'-\frac{\ell \left(8\pi r p -m''\right)}{\left(1+2\ell \right)8\pi r^2}.
\label{TOV}
\end{equation}
\textit{This is the modified TOV equation in the bumblebee context for a star that matches the Schwarzschild-like solution
for $r=R$, whose  matter content is described by a  perfect fluid.} 
The TOV equation, whether the original or the modified ones, regards the hydrostatic equilibrium inside the star. A given solution of the TOV provides, from a specific equation of state,  both the pressure and energy density configurations for a star.

From two components of the field equations, namely 
Eqs. (\ref{Eq00}) and (\ref{Eq11}), we obtain $\alpha'$, which is given by
\begin{equation}
\frac{d\alpha}{dr}=\frac{\left(1+2\ell\right)8 \pi  r^3 p +\left(2+ 3 \ell \right) m - \ell r m'}{\left(2+3 \ell\right) r (r-2 m)}.
\label{alpha}
\end{equation}
Therefore, any interior metric  should be calculated from Eqs. (\ref{TOV})-(\ref{alpha}) 
by using a specific equation of state. 
In this work, as mentioned before, we focus on the incompressible case and quark stars with the MIT bag 
equation of state.  

When the mass function $m(r) $ is previously specified, 
the (modified) TOV equation is an example of Riccati's equation, a nonlinear differential equation. 
Such an equation could be solved from a linear second-order 
differential equation.\footnote{See Ref. \cite{Ince}, chapter  2, for more details.}
 That is, if we write the modified TOV equation  (\ref{TOV}), with the aid of Eq. (\ref{alpha}),  as
\begin{equation}
\frac{dp}{dr}=q_0 + q_1 p +q_2 p^2,
\end{equation} 
it is possible to show that $p$ is solution of 
\begin{equation}
u''-q_3 u'+q_4u=0,
\label{u}
\end{equation}
in such a way that $q_0,q_1,q_2,q_3=q_1+q'_2/q_2,$ and $q_4=q_0q_2$ are functions of the radial coordinate (except for $q_2$,  their values will not be explicitly written here). 
The solution of (\ref{u})---in general large expressions in terms of 
hypergeometric functions---has two constants,
which are fixed by the boundary conditions: (i) the pressure is zero on the star surface, $p(R)=0$, and (ii) $\alpha(R)=\frac{1}{2}\ln \left(1-\frac{2M}{R} \right)$ is 
necessary to match the exterior solution.  
From the \textit{trick} of turning the nonlinear differential equation (\ref{TOV}) 
into a linear and second-order equation (\ref{u}), the pressure $p$ is simply 
\begin{equation}
p=-\frac{1}{q_2}\frac{u'}{u}=\frac{(2+3\ell)(r-2m)}{8\pi r^2 (1+2\ell)}\frac{u'}{u}.
\label{p_general}
\end{equation}
As we can see, with a given $p$ and $m$, Eq. (\ref{alpha}) provides the metric component $g_{tt}$. 
Assuming that $p$ comes
from Eq. (\ref{p_general}), the metric component is
\begin{equation}
\alpha=\int \frac{(2+3\ell)m -\ell r m'}{(2+3\ell)r(r-2m)}dr +\ln u, \ \ \ \mbox{for} \   r<R.
\label{gtt}
\end{equation}
It is worth emphasizing that the metric functions (\ref{grr}) and (\ref{gtt}) are exact solutions of the modified 
field equations (\ref{Eq00})-(\ref{Eq22}).
In order to be an interior solution, $e^{2\beta}$ and $e^{2\alpha}$ should be positive definite 
and should keep the metric signature inside the star, i.e.  for $r<R$. 
As we will see, that is the case for the two type of fluids presented here.

\section{Solving the modified TOV}

\subsection{Constant density star}

The simplest case to be considered here is the incompressible fluid. In such a case, the energy density is constant inside 
the star, that is, 
\begin{equation}
\rho(r)=\rho_0.
\end{equation}
With a constant density, Eq (\ref{mass}) gives us $m(R)=M=\frac{4}{3}\pi R^3\rho_0$ for the total mass of the star.
In the general relativity context, this simple case provides analytic expressions for the fluid pressure and the interior
metric. This is the case in the bumblebee context, but solutions are much larger. As an alternative, we could make 
some approximations in order to get more reasonable expressions. For example, excluding terms like
$\ell^n$ for $n>1$ and 
the second term on the right side of the modified TOV equation (\ref{TOV}) (because $\ell \rho_0(p/\rho_0-1) \ll 1$),
the function $u$, which is solution of Eq. (\ref{u}), reads
\begin{equation}
u(r)= \left(3-8 \pi r^2 \rho_0 \right)^{\frac{2-3\ell}{8}} \left(C_1 \left(3-8 \pi r^2 \rho_0 \right)^{\frac{1 }{2}}+C_2\right),
\label{u_approx}
\end{equation}
where the two constants $C_1$ and $C_2$ are fixed from the boundary conditions using Eqs. (\ref{alpha}) and (\ref{p_general}), which connect $u$, $p$, and, consequently, $\alpha$. For the approximation adopted, one has
\begin{align}
C_1 = & - \frac{\left(2- 3\ell \right)}{4\sqrt{3}} , \\
C_2 = & \frac{\sqrt{3}}{4}(2 -\ell)\left(3-8 \pi R^2 \rho_0 \right)^{\frac{1}{2}}.
\end{align} 
The general relativity values are restored for $\ell=0$. Within this approximation, the fluid pressure is given by
\begin{equation}
 p(r) = \rho_0 \left( \frac{\sqrt{1-\frac{2M}{R}}-\sqrt{1-\frac{2M r^2}{R^3}}}{\sqrt{1-\frac{2M r^2}{R^3}}-3\sqrt{1-\frac{2M}{R}}}\right) + \ell \rho_0 \left( \frac{ 7 - \frac{2M}{R} \left(\frac{r^2}{R^2} +6 \right) - 7 \sqrt{\left(1-\frac{2M r^2}{R^3}\right)\left(1-\frac{2M}{R} \right)} }{\left(\sqrt{1-\frac{2M r^2}{R^3}}-3\sqrt{1-\frac{2M}{R}} \right)^2}\right),
 \label{p_approx}
\end{equation}
which is the general relativity result plus a small contribution of the 
bumblebee field in this approximation. Notice also that the boundary condition $p(R)=0$ is satisfied.
 For $r=0$, the central pressure $p(0)=p_c$ diverges with $R=\frac{9}{4}M$, like the general relativity case.
That is, any reasonable star should have $R>\frac{9}{4}M$ or, equivalently, $M<\frac{4R}{9}$
(this is the so-called Buchdahl's limit in general relativity \cite{Buchdahl:1959zz}). 
Such a result is expected, 
for the bumblebee field does not change the event horizon radius in the exterior metric.

\begin{figure}
\begin{center}
\includegraphics[trim=0.27cm 0cm 0.7cm 0cm, clip=true,scale=0.55]{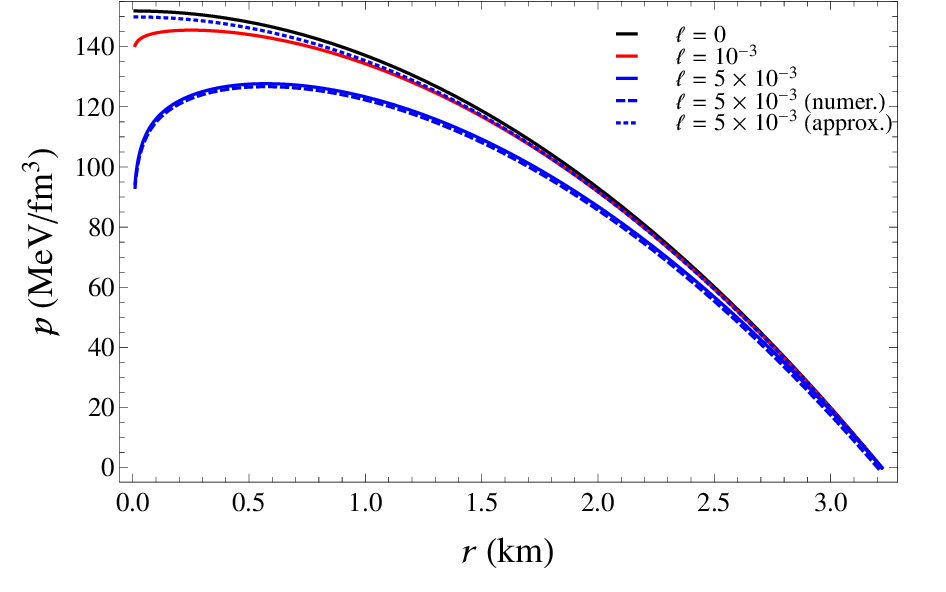}
\includegraphics[trim=0.08cm 0cm 0cm 0.03cm, clip=true,scale=0.55]{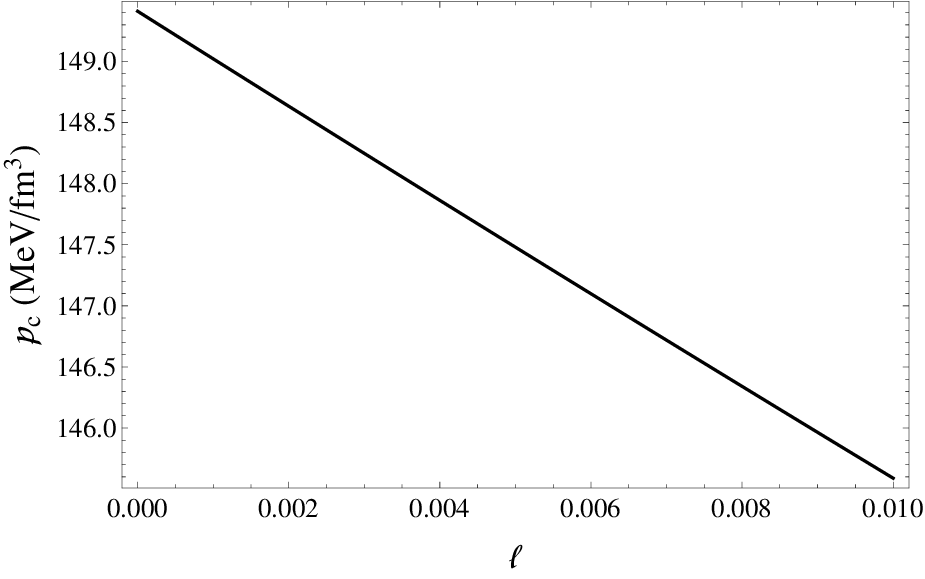}
\caption{Pressure (left) and central pressure (right) for a star with $\rho=\rho_0=2$ GeV/fm$^{3}$ 
and $R=3.2$ km. In the pressure graphic, exact, numerical, and approximate calculations are compared each other.
The bumblebee field decreases the central pressure of the star as $\ell$ increases.
However, the approximation breaks for $\ell > 10^{-3}$.}
\label{rho_constant}
\end{center}
\end{figure}

\begin{figure}
\begin{center}
\includegraphics[trim=0.35cm 0cm 0.7cm 0cm, clip=true,scale=0.57]{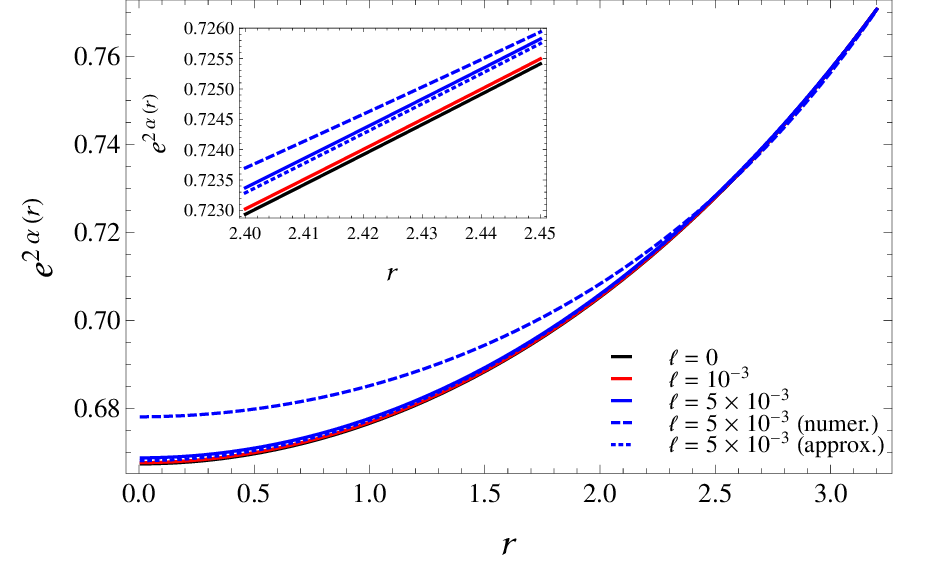}
\includegraphics[trim=0.35cm 0cm 0.7cm 0cm, clip=true,scale=0.575]{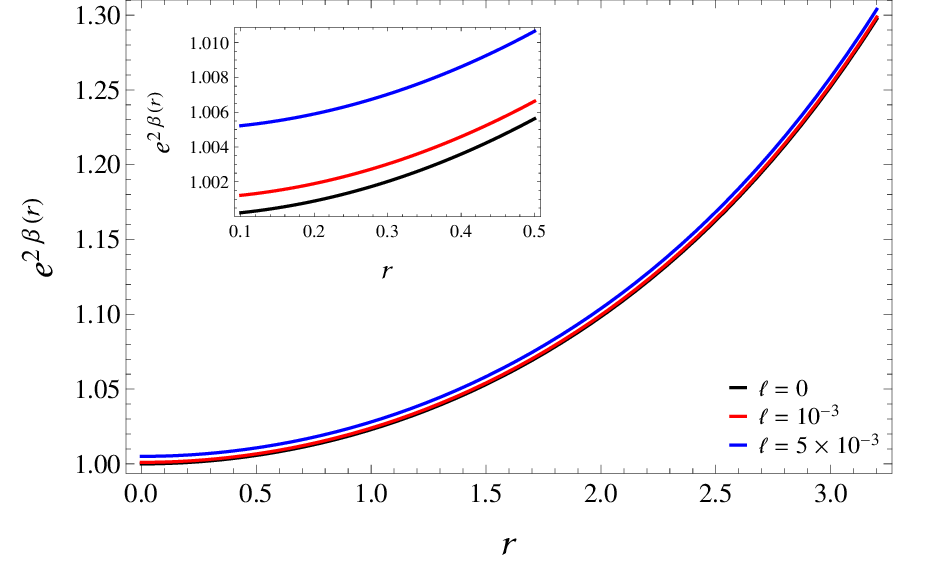}
\caption{Positivity of the metric functions, $g_{tt}=e^{2\alpha}$ and $g_{rr} =e^{2\beta}$, for very
exaggerated values of the Lorentz-violating parameter $\ell$ using a constant density fluid.
 Exact, numerical, and approximate calculations are compared each other. We adopt $R=3.2$ km and 
 $\rho=\rho_0=2$ GeV/fm$^{3}$.}
\label{rho_constant-metric}
\end{center}
\end{figure}

Fig. \ref{rho_constant} shows the star pressure (exactly, numerically, and approximately calculated) 
with the same radius $R=3.2$ km (thus same mass) for different 
values of the Lorentz-violating parameter $\ell$. As we can see,  the bumblebee field decreases 
the central pressure of the star.  This will no longer be the case for the quark matter case in the next section.
This effect in the constant-$\rho$ case is very counter-intuitive, because in this type of fluid the pressure 
decreases once the radius decreases. Using  the approximate pressure  (\ref{p_approx}) for $r=0$, 
it is possible to illustrate the relation between both the central pressure and the parameter $\ell$ in order to see
the effect of the bumblebee field on $p_c$. In the same graphic (right), we see that the
approximation (\ref{p_approx}) is no longer valid for $\ell >10^{-3}$, as we compare it with the pressure 
graphic (left). 

The interior metric (without any approximation) for constant energy density is straightforwardly 
calculated from Eq. (\ref{gtt}) and $m(r)=\frac{4}{3}\pi r^3 \rho_0$. That is,
\begin{equation}
\alpha (r) = \ln \left[\frac{u(r)}{\left(3-8 \pi r^2 \rho_0  \right)^{\frac{1}{4+6\ell}}} \right].
\end{equation}
Taking then the approximation for $u(r)$, given by Eq. (\ref{u_approx}), one has
\begin{equation}
e^{\alpha(r)}= \frac{3}{2}\sqrt{1-\frac{2M}{R}}-\frac{1}{2}\sqrt{1-\frac{2Mr^2}{R^3}}+\frac{3\ell}{4}\bigg( \sqrt{1-\frac{2Mr^2}{R^3}} -\sqrt{1-\frac{2M}{R}}\bigg).
\end{equation}
Even this approximation is in agreement with the boundary 
condition $\alpha(R)=\frac{1}{2}\ln \left(1-\frac{2M}{R} \right)$.
As we can see in Fig. \ref{rho_constant-metric}, for the exact, numerical, and approximate calculations, 
the metric components $e^{2\alpha}$ and 
$e^{2\beta}$ are positive definite inside the star. Thus, we have the correct metric signature both
inside and outside the star.

\subsection{Quark matter star}

In the second type of fluid studied, we assume a star in the bumblebee gravity made up of quark matter with an equation of state more \textit{realistic} than the one used in the constant-density star case. The quark matter could be described by the MIT bag model and by the following equation of state
\begin{equation}
p=\frac{1}{3} \left( \rho - 4 \mathcal{B}\right),
\label{MIT}
\end{equation}
where, as we mentioned, $\mathcal{B}$ is the bag constant. The MIT bag model confines quarks inside a bag, and the bag's energy, given as $E(r)=  \frac{4\pi}{3}r^3 \mathcal{B}+\frac{C}{r}$, combines a term proportional to 
the quarks' kinetic energy and their volume. The bag energy takes place when the energy 
reaches its minimum, consequently determining the constant $C$. In this context, the bag constant is defined as $\mathcal{B}=E_0/(4V_0)$, with $E_0$ representing the energy and $V_0$ denoting the volume of the bag. For a neutron, $m\approx E_0 \approx 1 \text{ GeV}$ with a radius of $r_0\approx 1 \text{ fm}$, resulting in $\mathcal{B}\approx 60 \text{ MeV/fm}^3$ (or $8 \times 10^{-5}$ km$^{-2}$ in
geometrized units) for numerical calculations. This value falls within the range of the bag constant, 
$\mathcal{B} \sim 57-92$ MeV/fm$^3$,
allowing for the reproduction of hadronic masses at low-energy regimes \cite{DeTar:1983rw}, and has been 
utilized in Refs. \cite{Gondek-Rosinska:2008zmv, Fraga:2013qra, Arbanil:2016wud}. However, given that the 
MIT bag model is a simple model, in extreme energy regimes the strong coupling constant for quark 
interactions is less than that one utilized in \cite{DeTar:1983rw}. In extreme density regimes, 
the strong coupling constant is less than in typical conditions where values of $\mathcal{B}$ were derived.
 In this sense, it is justifiable to slightly adjust the bag constant.
 Thus, like Ref. \cite{Gondek-Rosinska:2008zmv}, we also adopt $\mathcal{B}=40$ MeV/fm$^3$ (equivalent to $5 \times 10^{-5}$ km$^{-2}$ in geometrized units). Unfortunately, for the equation of state (\ref{MIT}) analytic solutions are not possible.
 
\begin{figure}
\begin{center}
\includegraphics[trim=0.4cm 0cm 0.7cm 0cm, clip=true,scale=0.55]{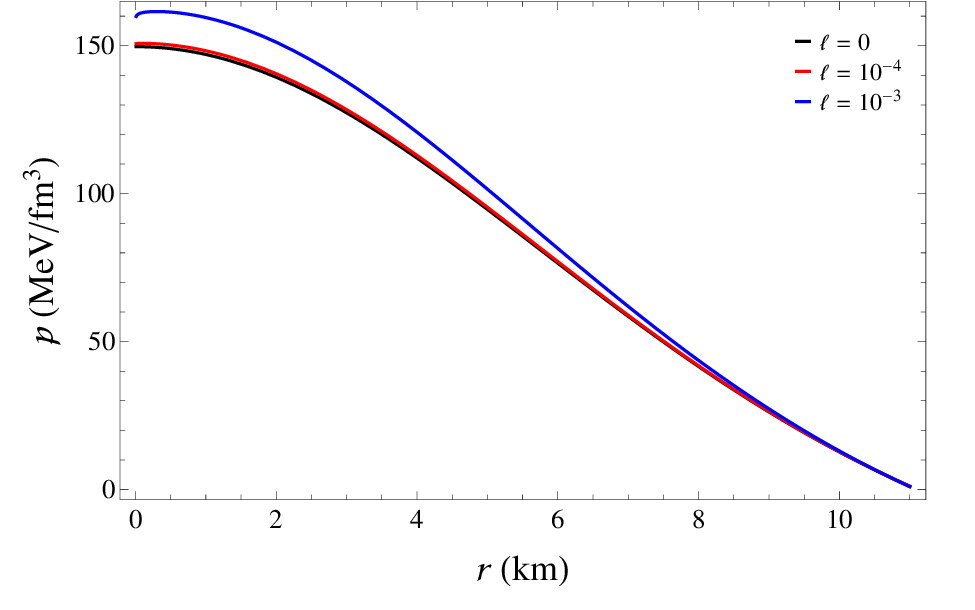}
\includegraphics[trim=0.4cm 0cm 0.7cm 0cm, clip=true,scale=0.55]{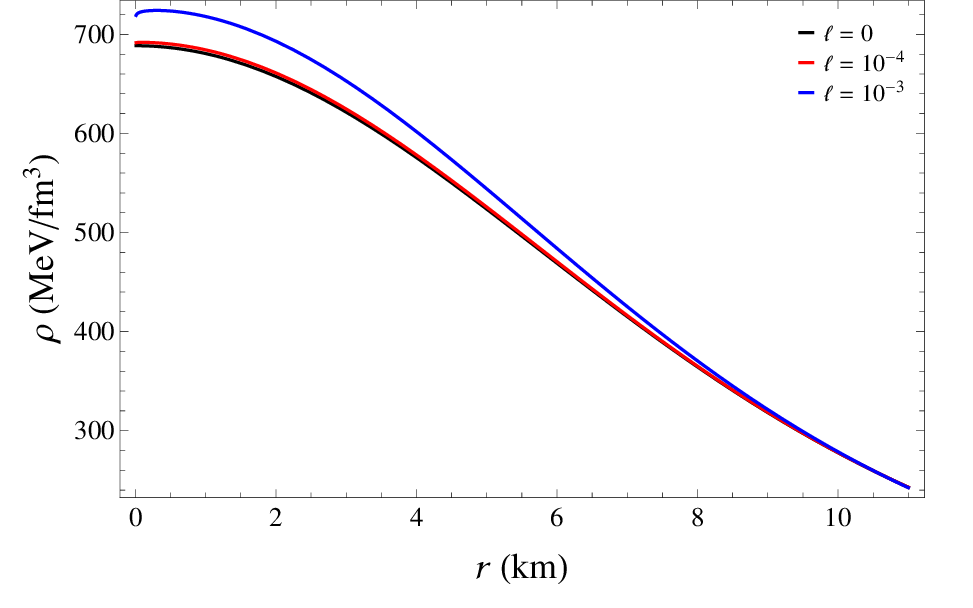}
\includegraphics[trim=0.4cm 0cm 0.7cm 0cm, clip=true,scale=0.55]{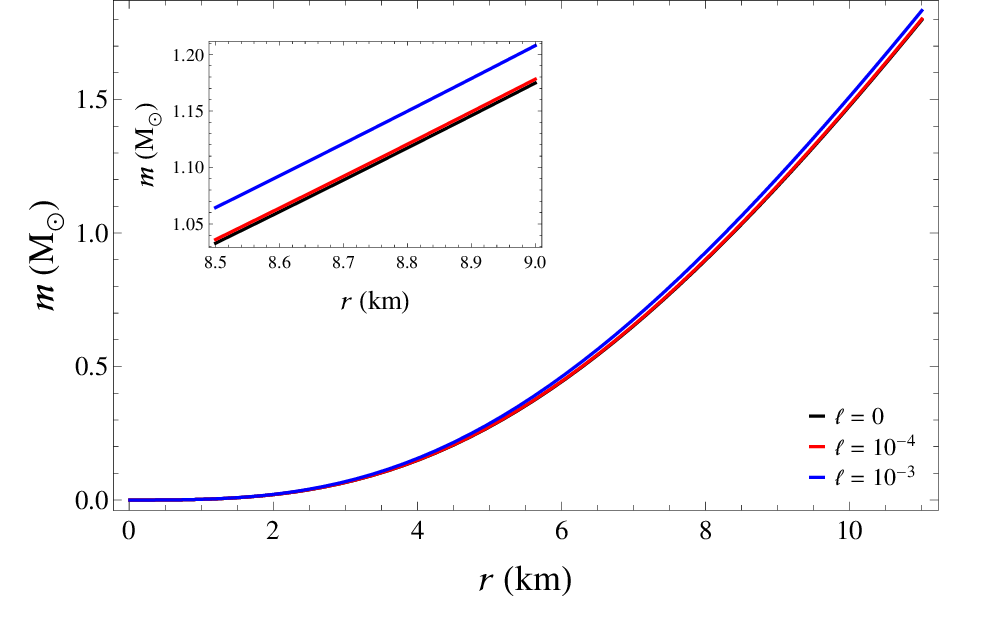}
\caption{Pressure (left), energy density (right) and mass function (bottom) with the MIT bag equation of state. 
The bumblebee field, represented by the Lorentz-violating parameter $\ell$, increases the three quantities. We adopted
$\mathcal{B}=60$ MeV/fm$^{3}$ and $R=11$ km in these graphics.}
\label{p-rho-MIT}
\end{center}
\end{figure}
 
\begin{figure}[!ht]
\begin{center}
\includegraphics[trim=0.35cm 0cm 0.7cm 0cm, clip=true,scale=0.59]{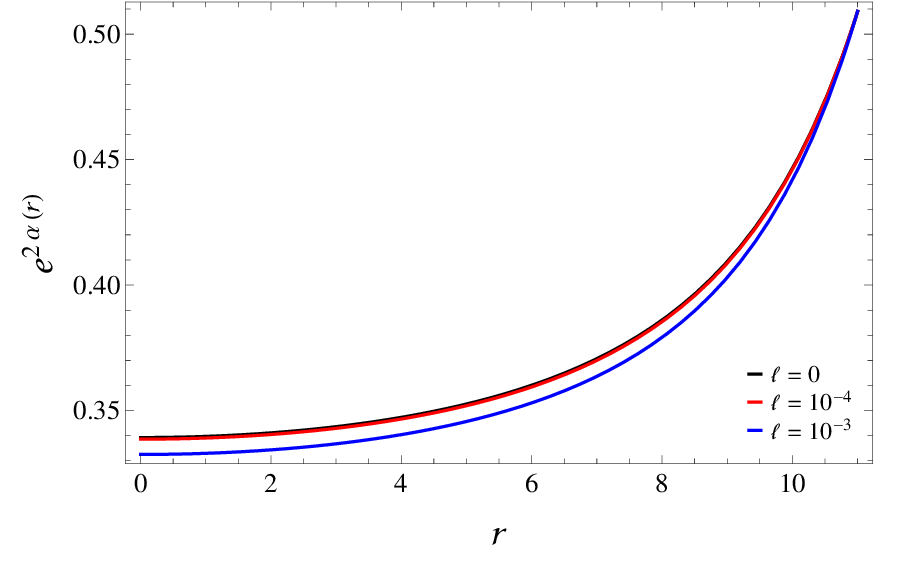}
\includegraphics[trim=0.3cm 0cm 0.9cm 0cm, clip=true,scale=0.585]{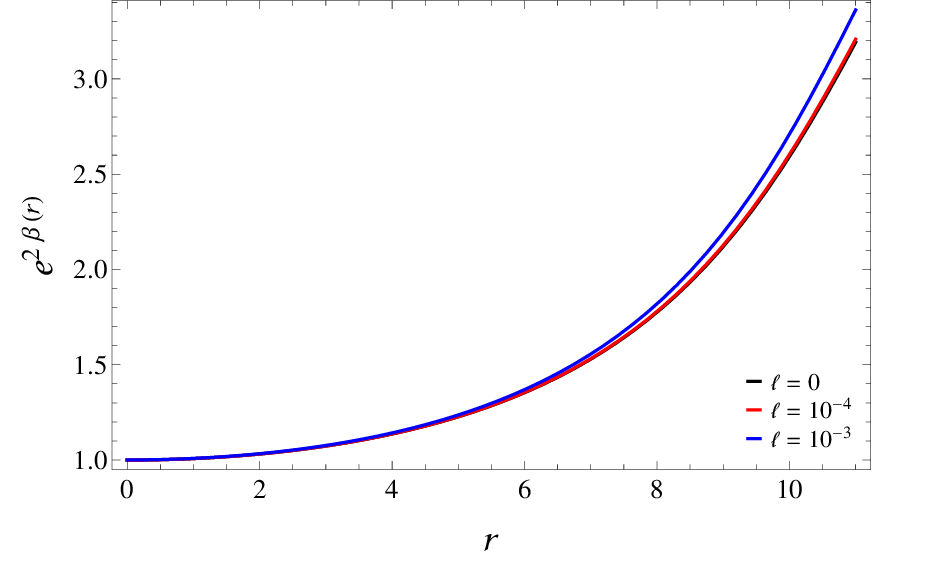}
\caption{Positivity of the metric functions, $g_{tt}=e^{2\alpha}$ and $g_{rr} =e^{2\beta}$, for very
exaggerated values of the Lorentz-violating parameter $\ell$ using the MIT bag model. We adopt $R=11$ km in
 these graphics.}
\label{MIT-metric}
\end{center}
\end{figure} 
 
Once again, according to Fig. \ref{p-rho-MIT}, 
like the constant density case, the bumblebee field modifies all physical parameters.
Interestingly, contrary to the constant density fluid, given a star with same radius, the bumblebee model leads to 
higher central pressures when compared to the general relativity predictions. For example, for 
a star with $R=11.1$ km, using the MIT bag 
equation of state with $\mathcal{B}= 60$ MeV/fm${^3}$,
 general relativity predicts $p_c= 150$ Mev/fm$^{3}$ for 
the central pressure. On the other hand, for $\ell = 10^{-2}$, the bumblebee model gives us
$p_c= 160$ MeV/fm$^{3}$. Consequently, the energy density also increases as the bumblebee field
reaches the VEV. In this case, the explanation comes from the mass function $m(r)$, which also increases
 due to bumblebee field acting on the spacetime geometry (see Fig. \ref{p-rho-MIT} on the bottom).

Like the constant-density case, Fig. \ref{MIT-metric} shows the metric functions $e^{\alpha}$
 and $e^{\beta}$ are positive definite inside the star $(r<R)$, consequently
the metric is well behaved and its signature $(-,+,+,+)$ is also preserved.  

\begin{figure}
\begin{center}
\includegraphics[trim=0.6cm 0cm 0.9cm 0cm, clip=true,scale=0.36]{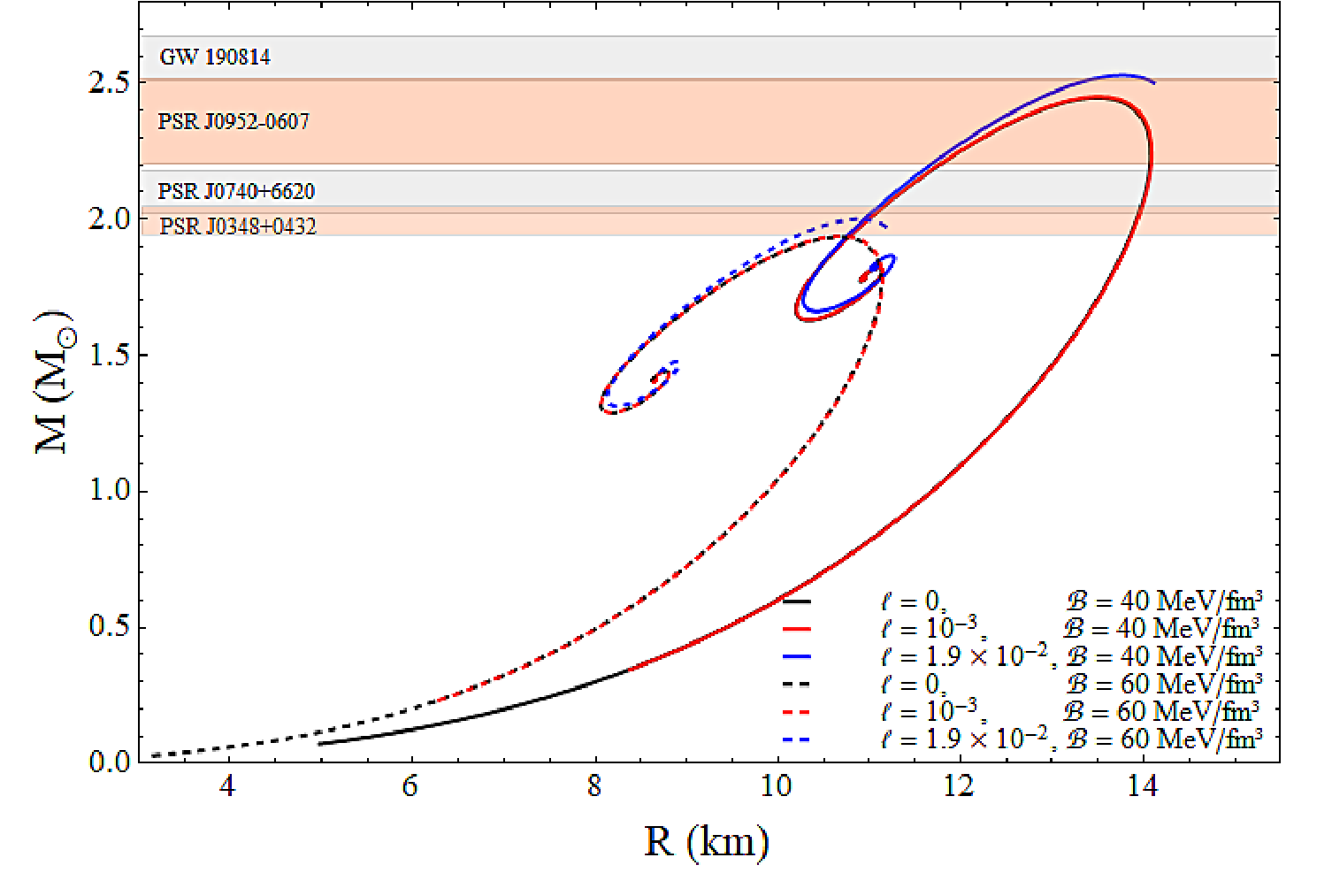}
\includegraphics[trim=0cm 0cm 0.7cm 0cm, clip=true,scale=0.495]{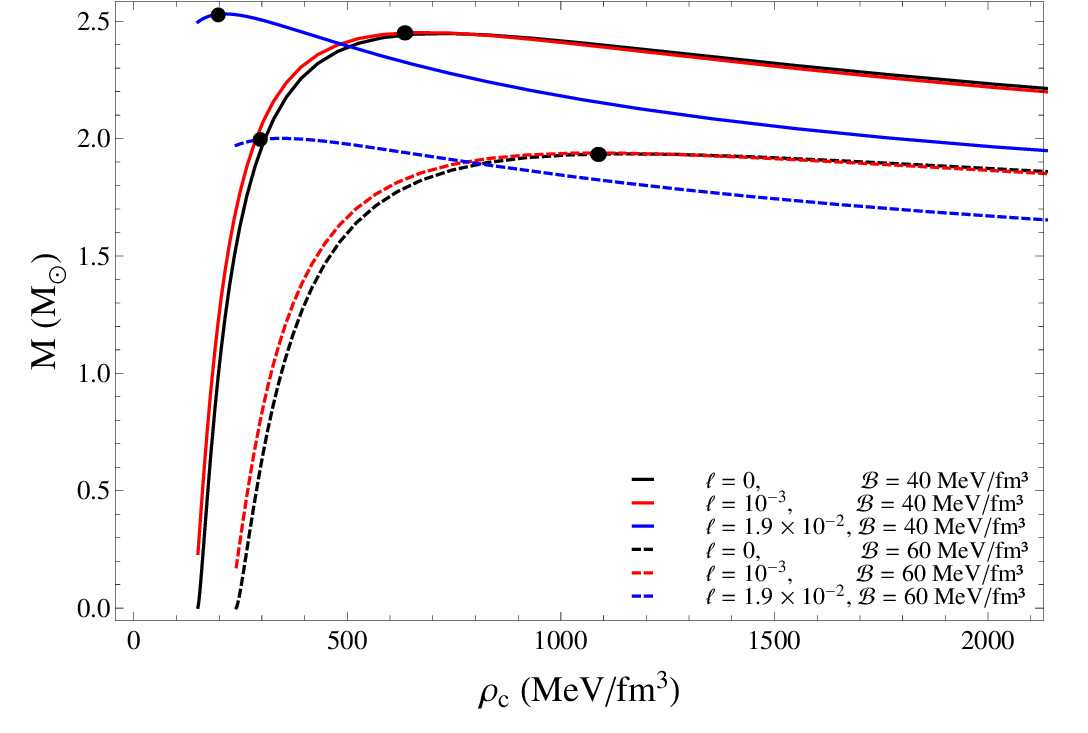}
\caption{On the left, the mass-radius relation in the bumblebee context for quark matter described by the MIT bag model
with two values for the bag constant $\mathcal{B}$. For large values of the Lorentz-violating
parameter $\ell$, the bumblebee gravity could provide larger mass-radius relations than general
relativity. The pulsars PSR J0348+0432 \cite{Antoniadis:2013pzd} $(M=2.01 \pm 0.04 M_{\odot})$, 
PSR J0740+6620 \cite{Fonseca:2021wxt} 
$(M=2.08 \pm 0.07 M_{\odot})$, and PSR J0952-0607 \cite{Romani:2022jhd} $(M=2.35 \pm 0.17 M_{\odot})$ 
are described/predicted by general relativity
with the bag constant $\mathcal{B}=40$ MeV/fm$^{3}$. On the other hand, the compact star in the 
GW 190814 \cite{LIGO} event, with $M=2.59^{-0.08}_{-0.09} M_{\odot}$, is not.
But it could be in the bumblebee context. See Table \ref{Parameters}
for a more detailed comparison. On the right, the $M-\rho_c$ curve shows 
stability just for $\frac{dM}{d\rho_c}>0$. 
The black dots $(\frac{dM}{d\rho_c}=0)$ indicate the limit of stable configurations.}
\label{MR}
\end{center}
\end{figure}

As mentioned in Introduction, the bumblebee model could provide more massive stars than the 
general relativity context, for the mass-radius relation is also modified due to the bumblebee field and its VEV.
As we can see in Fig. \ref{MR}, large values for the Lorentz-violating parameter $\ell$ increases the mass-radius
relation dramatically, except for a threshold (about $\ell = 10^{-1}$) from which larger values of $\ell$
 do not increase that relation. Also, it is worth mentioning the \enquote{incomplete} curves for large
 values of $\ell$ in Fig. \ref{MR}. 
 A given curve in the mass-radius graphic is solution of the TOV with conditions like
 $p(r) \geq 0$ and $r-2m(r) \geq 0$. For large values of the Lorentz-violating parameter, those conditions
 are satisfied just for smaller range of parameters. As we will see, this is linked to the stability of 
 the solutions.

 Therefore, for appropriate values of the Lorentz-violating parameter $\ell$, 
 a modified gravitational model like the bumblebee model 
could fit data for large mass-radius relations in which the general relativity context is not able to fit. 
In particular, we see in Fig. \ref{MR} and Table \ref{Parameters} that for $\mathcal{B}=60$ MeV/fm$^3$ 
and $\ell=  1.9 \times 10^{-2}$, the 
bumblebee model give us a star with $R=2  M_{\odot}$. On the other hand, general relativity is not able to
reach the  $R=2 M_{\odot}$ limit. With a small bag constant (as we can also see  in Table \ref{Parameters}), 
 for example $\mathcal{B}=40$ MeV/fm$^3$,
the bumblebee model is able to describe even more massive stars, like the one
in the gravitational wave event GW 190814, with mass $M=2.59^{-0.08}_{-0.09} M_{\odot}$. Contrary
to general relativity, which does not fit this data ($M=2.45 M_{\odot}$), the bumblebee model with 
$\ell= 1.9 \times 10^{-2}$ predicts a massive quark star with $M=2.53 M_{\odot}$ and $R=13.75$ km. 
This is a virtue of modified theories of gravity \cite{Gammon:2023uss,Tangphati:2024war,Haghani:2023nrm}. 
Because theories that predict smaller mass-radius relations than general relativity does 
would be under pressure due to recent data.

 \begin{table}
\begin{ruledtabular}
\begin{tabular}{lccc}
     
       \\  $\mathcal{B}=40$ MeV/fm$^{3}$ & $\ell = 0$ & $\ell = 10^{-3}$ & $\ell =1.9 \times 10^{-2}$  \\  \\ \hline
      $M(M_{\odot})$ & 2.45 & 2.45 & 2.53 \\ 
      $R$ (km) & 13.51 & 13.54 & 13.75 \\ 
      $p_c$ (MeV/fm$^{3}$) & 187.12 & 172.15 & 22.45 \\ 
       $\rho_c$ (MeV/fm$^{3}$) & 711.05 & 666.15 & 217.06 \\  \hline \\
      
       $\mathcal{B}=60$ MeV/fm$^{3}$ & & &   \\ \\ \hline
      $M(M_{\odot})$ & 1.94 & 1.94 & 2.0 \\ 
      $R$ (km) & 10.66 & 10.68 & 10.87 \\ 
      $p_c$ (MeV/fm$^{3}$) & 306.88 & 284.42 & 37.42 \\ 
       $\rho_c$ (MeV/fm$^{3}$) & 1,070.32 & 1,002.96 & 261.97 \\

\end{tabular}
\end{ruledtabular}
\caption{Star parameters for values of the bag constant $\mathcal{B}$ and symmetry breaking 
parameter $\ell$. $M$ is the mass of the star, $R$ is its radius, $p_c$ is the central pressure,  and $\rho_c$ 
is the central density.}
\label{Parameters}

\end{table}

\section{Stability analysis}
In this section, we briefly comment stability of the bumblebee star with quark matter
described by the MIT bag equation of state (\ref{MIT}). In general relativity, 
a necessary but not sufficient stability criterion regards
the central pressure $\rho_c$. It is assumed, even in modified gravity works, that $\frac{dM}{d\rho_c}>0$
is a necessary condition to stability in the static case \cite{Zeldovich,Olmo:2019flu}. 
As we can see in Fig. \ref{MR}, the $M-\rho_c$ curve indicates
 stability intervals for both different values of the bag constant $\mathcal{B}$ and the
Lorentz-violating parameter $\ell$. Also we can see that large values of $\ell$ yield small
stable ranges. This is in agreement with the mass-radius diagram that shows \enquote{incomplete}
curves for large values of $\ell$. 

It is common to adopt the speed of sound in the fluid, 
$c_s =\sqrt{\frac{dp}{d\rho}}$, to check whether or not causality and stability issues could arise. 
Inside a stable star, the speed of sound of the fluid must never surpass the speed of light $c$. Using the MIT
bag equation of state, it is easy to see that $c_s = \frac{c}{\sqrt{3}}$ (where the speed of light in vacuum was restored), 
that is, one has  a constant subluminal speed of sound in the fluid, thus no causality and stability issues are present
due to the speed of sound. 

Another useful parameter to investigate the star stability is the adiabatic index of perturbations, namely
\begin{equation}
\gamma=\left(1+\frac{\rho}{p} \right) \left(\frac{dp}{d\rho} \right)_{S},
\end{equation} 
where $S$ means that the derivative (or the speed of sound in the fluid) 
is calculated at a constant specific entropy. In the general relativity context, for spherical symmetry, 
Chandrasekhar \cite{Chandrasekhar:1964zz}
showed using radial perturbations that $\gamma_{cr} = \frac{4}{3}+\frac{19}{42}b$ is the critical value of
the adiabatic index for a given stable star, where $b$ is the compactness parameter. 
That is, stable stars must have $\gamma>\gamma_{cr}$.
In bumblebee gravity (and other modified
gravity theories) there is no available  $\gamma_{cr}$ yet. Thus, it is common to assume
that stable stars in these alternative contexts must satisfy at least  $\gamma > \frac{4}{3}$, which
is the Newtonian limit for the adiabatic index.
This is the case here, for using the MIT bag equation of state (\ref{MIT}), one has 
$\gamma=\frac{1}{3}\left(4+\frac{1}{p}\right) > \frac{4}{3}$,  
according to the positivity of $p$ in Fig. \ref{p-rho-MIT}, suggesting then stability.

\section{Final remarks}
The bumblebee gravity is a Lorentz-violating model in which the VEV of the bumblebee field is nonzero, modifying
then fields and spacetime. Within this context of modified theory of gravity, spherically symmetry metrics, describing
the interior of noncharged and nonrotating stars, were obtained analytically. 

By using two types of fluids, we show that the bumblebee field could increase or decrease the fluid 
pressure inside the star. In particular for the more \textit{realistic}  fluid studied in this article, 
quark matter described by the MIT bag model, the Lorentz violation increases both the pressure and the 
energy density of the fluid of quark stars with same radius. 

In the end of the day, it is reasonable to hope for the Lorentz breaking symmetry in contexts of high energy and 
intense gravitational effects. The results presented in this article point out to larger mass-radius
relations in the bumblebee context when compared to the general relativity case. In this sense, recent data of large
mass-radius relations, like the gravitational wave events, GW 190814, 
with mass $M=2.59^{-0.08}_{-0.09} M_{\odot}$, could be justified by a modified gravity model like the
bumblebee model using quark matter as the star matter content.

Another interesting point regards the stability analysis. The $M - \rho_c$ 
curve---from the $\frac{dM}{d\rho_c}>0$ stability criterion---indicates 
stability for the bumblebee stars with quark matter. Moreover, large values of the 
Lorentz-violating parameter $\ell$ provide larger mass-radius relations, but the 
range of allowed and stable configurations decreases with that parameter.

\section*{Acknowledgments}
JCSN thanks Conselho Nacional de Desenvolvimento Científico e Tecnológico (CNPq), Brazil, 
(Grant No. 170579/2023-9) for the financial support and the ICT-Unifal for the kind hospitality. 
F.G.G. is supported by CNPq through 306762/2021-8. The authors also thank two anonymous referees
for valuable comments.



\begin{thebibliography}{99}


\bibitem{Liberati2013}
S.~Liberati, 
Tests of Lorentz invariance: a 2013 update,
Class. Quant. Grav. \textbf{30} (2013) 133001, arXiv:1304.5795. 


\bibitem{Berti:2015itd}
E.~Berti, E.~Barausse, V.~Cardoso, L.~Gualtieri, P.~Pani, U.~Sperhake, L.~C.~Stein, N.~Wex, K.~Yagi and T.~Baker, \textit{et al.}
Testing General Relativity with Present and Future Astrophysical Observations,
Class. Quant. Grav. \textbf{32}  (2015) 243001,  arXiv:1501.07274.

\bibitem{Addazi:2021xuf}
A.~Addazi, J.~Alvarez-Muniz, R.~Alves Batista, G.~Amelino-Camelia, V.~Antonelli, M.~Arzano, M.~Asorey, J.~L. Atteia, S.~Bahamonde and F.~Bajardi,  \textit{et al.}
Quantum gravity phenomenology at the dawn of the multi-messenger era---A review,
Prog. Part. Nucl. Phys. \textbf{125} (2022) 103948,  arXiv:2111.05659. 

\bibitem{Berti:2018cxi}
E.~Berti, K.~Yagi and N.~Yunes,
Extreme Gravity Tests with Gravitational Waves from Compact Binary Coalescences: (I) Inspiral-Merger,
Gen. Rel. Grav. \textbf{50} (2018) 46, arXiv:1801.03208.

\bibitem{Kostelecky1989} 
V.~A.~Kostelecky and S.~Samuel,
Spontaneous Breaking of Lorentz Symmetry in String Theory,
Phys. Rev. D \textbf{39} (1989) 683.

\bibitem{Kostelecky1989a}
V.~A.~Kostelecky and S.~Samuel,
Phenomenological Gravitational Constraints on Strings and Higher Dimensional Theories,
Phys. Rev. Lett. \textbf{63} (1989) 224.

\bibitem{Kostelecky1989b}
V.~A.~Kostelecky and S.~Samuel,
Gravitational Phenomenology in Higher Dimensional Theories and Strings,
Phys. Rev. D \textbf{40}  (1989) 1886.

\bibitem{Kostelecky1991}
V.~A.~Kostelecky and R.~Potting,
CPT and strings,
Nucl. Phys. B \textbf{359}  (1991) 545.

\bibitem{Santos:2012if}
V.~Santos and C.~A.~S.~Almeida,
On Gravity localization under Lorentz Violation in warped scenario,
Phys. Lett. B \textbf{718}  (2013) 1114, arXiv:1211.4542.

\bibitem{Carroll:2001ws}
S.~M.~Carroll, J.~A.~Harvey, V.~A.~Kostelecky, C.~D.~Lane and T.~Okamoto,
Noncommutative field theory and Lorentz violation,
Phys. Rev. Lett. \textbf{87} (2001) 141601, arXiv:hep-th/0105082.

\bibitem{Horava2009} 
P.~Horava,
Quantum Gravity at a Lifshitz Point,
Phys. Rev. D \textbf{79} (2009) 084008,  arXiv:0901.3775.

\bibitem{Kostelecky1997} 
D.~Colladay and V.~A.~Kostelecky,
CPT violation and the standard model,
Phys. Rev. D \textbf{55} (1997) 6760, arXiv:hep-ph/9703464.

\bibitem{Kostelecky1998}
D.~Colladay and V.~A.~Kostelecky,
Lorentz violating extension of the standard model,
Phys. Rev. D \textbf{58} (1998) 116002, arXiv:hep-ph/9809521.

\bibitem{Bluhm:1997qb}
R.~Bluhm, V.~A.~Kostelecky and N.~Russell,
CPT and Lorentz tests in Penning traps,
Phys. Rev. D \textbf{57} (1998) 3932, arXiv:hep-ph/9809543. 

\bibitem{Bluhm:1997qbc}
R.~Bluhm, V.~A.~Kostelecky and N.~Russell, 
CPT and Lorentz tests in hydrogen and anti-hydrogen,
Phys. Rev. Lett. \textbf{82} (1999) 2254, arXiv:hep-ph/9810269. 

\bibitem{Kostelecky:1999mr}
V.~A.~Kostelecky and C.~D.~Lane,
Constraints on Lorentz violation from clock comparison experiments,
Phys. Rev. D \textbf{60} (1999) 116010, arXiv:hep-ph/9908504. 

\bibitem{Bluhm:1999dx}
R.~Bluhm, V.~A.~Kostelecky and C.~D.~Lane,
CPT and Lorentz tests with muons,
Phys. Rev. Lett. \textbf{84} (2000) 1098, arXiv:hep-ph/9912451. 

\bibitem{Kostelecky:2000mm}
V.~A.~Kostelecky and R.~Lehnert,
Stability, causality, and Lorentz and CPT violation,
Phys. Rev. D \textbf{63} (2001) 065008, arXiv:hep-th/0012060. 

\bibitem{Colladay:2001wk}
D.~Colladay and V.~A.~Kostelecky, 
Cross-sections and Lorentz violation,
Phys. Lett. B \textbf{511}  (2001) 209, arXiv:hep-ph/0104300.

\bibitem{Altschul:2004xp}
B.~Altschul,
Compton scattering in the presence of Lorentz and CPT violation,
Phys. Rev. D \textbf{70} (2004) 056005, arXiv:hep-ph/0405084. 

\bibitem{Carroll:1989vb}
S.~M.~Carroll, G.~B.~Field and R.~Jackiw,
Limits on a Lorentz and Parity Violating Modification of Electrodynamics,
Phys. Rev. D \textbf{41} (1990) 1231.

\bibitem{Andrianov:1998wj}
A.~A.~Andrianov and R.~Soldati,
Patterns of Lorentz symmetry breaking in QED by CPT odd interaction,
Phys. Lett. B \textbf{435} (1998) 449, arXiv:hep-ph/9804448. 

\bibitem{Araujo:2023izx}
M.~C.~Ara\'ujo, J.~Furtado and R.~V.~Maluf,
Lorentz-violating extension of scalar QED at finite temperature,
Phys. Lett. B \textbf{844} (2023) 138064, arXiv:2306.06959.

\bibitem{Jackiw:1999yp}
R.~Jackiw and V.~A.~Kostelecky,
Radiatively induced Lorentz and CPT violation in electrodynamics,
Phys. Rev. Lett. \textbf{82}  (1999) 3572, arXiv:hep-ph/9901358. 

\bibitem{Perez-Victoria:1999erb}
M.~Perez-Victoria,
Exact calculation of the radiatively induced Lorentz and CPT violation in QED,
Phys. Rev. Lett. \textbf{83} (1999) 2518, arXiv:hep-th/9905061. 

\bibitem{Altschul:2004gs}
B.~Altschul,
Gauge invariance and the Pauli-Villars regulator in Lorentz- and CPT-violating electrodynamics,
Phys. Rev. D \textbf{70} (2004) 101701, arXiv:hep-th/0407172. 

\bibitem{Nascimento:2007rb}
J.R.~Nascimento, E.~Passos, A.Y.~Petrov and F.A.~Brito,
Lorentz-CPT violation, radiative corrections and finite temperature,
JHEP \textbf{06}  (2007) 016, arXiv:0705.1338. 

\bibitem{Belchior:2023cbl}
F.~M.~Belchior and R.~V.~Maluf,
One-loop radiative corrections in bumblebee-Stueckelberg model,
Phys. Lett. B \textbf{844} (2023) 138107, arXiv:2307.14252. 

\bibitem{Bluhm2005}
R.~Bluhm and V.~A.~Kostelecky,
Spontaneous Lorentz violation, Nambu-Goldstone modes, and gravity,
Phys. Rev. D \textbf{71}  (2005) 065008, arXiv:hep-th/0412320. 

\bibitem{Bluhm2008}
R.~Bluhm, S.~H.~Fung and V.~A.~Kostelecky,
Spontaneous Lorentz and Diffeomorphism Violation, Massive Modes, and Gravity,
Phys. Rev. D \textbf{77} (2008) 065020, arXiv:0712.4119. 

\bibitem{Kostelecky2004} 
V. ~Kostelecky,
Gravity, Lorentz violation, and the standard model,
Phys. Rev. D \textbf{69} (2004) 105009, arXiv:hep-th/0312310. 

\bibitem{Bailey2006} 
Q.~G.~Bailey and V.~A.~Kostelecky,
Signals for Lorentz violation in post-Newtonian gravity,
Phys. Rev. D \textbf{74} (2006) 045001, arXiv:gr-qc/0603030. 

\bibitem{Filho:2022yrk}
A.~A.~A.~Filho, J.~R.~Nascimento, A.~Y.~Petrov and P.~J.~Porf\'\i{}rio,
Vacuum solution within a metric-affine bumblebee gravity,
Phys. Rev. D \textbf{108} (2023) 085010, arXiv:2211.11821. 

\bibitem{Maluf:2014dpa}
R.~V.~Maluf, C.~A.~S.~Almeida, R.~Casana and M.~M.~Ferreira, Jr.,
Einstein-Hilbert graviton modes modified by the Lorentz-violating bumblebee Field,
Phys. Rev. D \textbf{90} (2014) 025007, arXiv:1402.3554.

\bibitem{Kostelecky:2016kkn}
V.  Kosteleck\'y, A. Melissinos and M. Mewes, 
Searching for photon-sector Lorentz violation using gravitational-wave  detectors,
Phys. Lett. B \textbf{761} (2016) 1,  arXiv:1608.02592.

\bibitem{Kostelecky:2016kfm}
V.~A.~Kosteleck\'y and M.~Mewes,
Testing local Lorentz invariance with gravitational waves,
Phys. Lett. B \textbf{757} (2016) 510, arXiv:1602.04782. 

\bibitem{Amarilo:2023wpn}
K.M.~Amarilo, M.B.F.~Filho, A.A.A.~Filho and J.~Reis,
Gravitational waves effects in a Lorentz-violating scenario,
Phys.Lett.B \textbf{855} (2024) 138785, arXiv:2307.10937. 

\bibitem{Euclides2020}
L.A.~Lessa, J.E.G.~Silva, R.V.~Maluf and C.A.S.~Almeida,
Modified black hole solution with a background Kalb--Ramond field,
Eur. Phys. J. C \textbf{80} (2020) 335,  arXiv:1911.10296.

\bibitem{Casana}
R.~Casana, A.~Cavalcante, F.~P.~Poulis and E.~B.~Santos,
Exact Schwarzschild-like solution in a bumblebee gravity model,
Phys. Rev. D \textbf{97} (2018) 104001, arXiv:1711.02273. 

\bibitem{Maluf_Neves}
R.~V.~Maluf and J.~C.~S.~Neves,
Black holes with a cosmological constant in bumblebee gravity,
Phys. Rev. D \textbf{103}  (2021) 044002, arXiv:2011.12841. 

\bibitem{Bertolami}
O.~Bertolami and J.~Páramos,
The Flight of the bumblebee: Vacuum solutions of a gravity model with vector-induced spontaneous Lorentz symmetry breaking,
Phys. Rev. D \textbf{72}  (2005) 044001, arXiv:hep-th/0504215. 

\bibitem{Rondineli2019} 
R. Oliveira, D. M. Dantas, V. Santos and C. A. S. Almeida,
Quasinormal modes of bumblebee wormhole,
Class. Quant. Grav. \textbf{36} (2019)  105013,  arXiv:1812.01798. 

\bibitem{Ovgun2020} 
A.~\"Ovg\"un, K.~Jusufi and \.I.~Sakall\i{}, 
Exact traversable wormhole solution in bumblebee gravity,
Phys. Rev. D \textbf{99} (2019) 024042, arXiv:1804.09911. 

\bibitem{Capelo:2015ipa}
D.~Capelo and J.~P\'aramos,
Cosmological implications of Bumblebee vector models,
Phys. Rev. D \textbf{91} (2015) 104007, arXiv:1501.07685. 

\bibitem{Petrov}
A.~F.~Santos, A.~Y.~Petrov, W.~D.~R.~Jesus and J.~R.~Nascimento,
G\"odel solution in the bumblebee gravity,
Mod. Phys. Lett. A \textbf{30}  (2015) 1550011, arXiv:1407.5985. 

\bibitem{ONeal}
K.~O'Neal-Ault, Q.~G.~Bailey and N.~A.~Nilsson,
3+1 formulation of the standard model extension gravity sector,
Phys. Rev. D \textbf{103} (2021) 044010, arXiv:2009.00949. 

\bibitem{Maluf:2021lwh}
R.~V.~Maluf and J.~C.~S.~Neves, 
Bumblebee field as a source of cosmological anisotropies,
JCAP \textbf{10}  (2021) 038, arXiv:2105.08659. 

\bibitem{Maluf:2021eyu}
R.~V.~Maluf and J.~C.~S.~Neves,
Bianchi type I cosmology with a Kalb\textendash{}Ramond background field,
Eur. Phys. J. C \textbf{82} (2022) 135, arXiv:2111.13165. 

\bibitem{Neves:2022qyb}
J.~C.~S.~Neves,
Kasner cosmology in bumblebee gravity,
Annals Phys. \textbf{454} (2023) 169338, arXiv:2209.00589. 

\bibitem{Gardim:2019xjs}
F.~G.~Gardim, G.~Giacalone, M.~Luzum and J.~Y.~Ollitrault,
Thermodynamics of hot strong-interaction matter from ultrarelativistic nuclear collisions,
Nature Phys. \textbf{16}  (2020) 615-619, arXiv:1908.09728.

\bibitem{Annala:2019puf}
E.~Annala, T.~Gorda, A.~Kurkela, J.~N\"attil\"a and A. Vuorinen,
Evidence for quark-matter cores in massive neutron stars,
Nature Phys. \textbf{16} (2020) 907, arXiv:1903.09121.

\bibitem{Chodos:1974je}
A.~Chodos, R.~L.~Jaffe, K.~Johnson, C.~B.~Thorn and V.~F.~Weisskopf,
A New Extended Model of Hadrons,
Phys. Rev. D \textbf{9} (1974) 3471.


\bibitem{Itoh:1970uw}
N.~Itoh,
Hydrostatic Equilibrium of Hypothetical Quark Stars,
Prog. Theor. Phys. \textbf{44} (1970) 291.

\bibitem{Farhi:1984qu}
E.~Farhi and R.~L.~Jaffe,
Strange Matter,
Phys. Rev. D \textbf{30} (1984) 2379.

\bibitem{Witten:1984rs}
E.~Witten,
Cosmic Separation of Phases,
Phys. Rev. D \textbf{30}  (1984) 272.

\bibitem{Fraga:2013qra}
E.~S.~Fraga, A.~Kurkela and A.~Vuorinen,
Interacting quark matter equation of state for compact stars,
Astrophys. J. Lett. \textbf{781} (2014) 2, L25, arXiv:1311.5154. 

\bibitem{Tangphati:2021tcy}
T.~Tangphati, A.~Pradhan, A.~Errehymy and A.~Banerjee,
Anisotropic quark stars in Einstein-Gauss-Bonnet theory,
Phys. Lett. B \textbf{819} (2021) 136423.

\bibitem{Tangphati:2021mvu}
T.~Tangphati, A.~Pradhan, A.~Errehymy and A.~Banerjee,
Quark stars in the Einstein\textendash{}Gauss\textendash{}Bonnet theory: A new branch of stellar configurations,
Annals Phys. \textbf{430} (2021) 168498.

\bibitem{Pretel:2021czp}
J.~M.~Z.~Pretel, A.~Banerjee and A.~Pradhan,
Electrically charged quark stars in 4D Einstein\textendash{}Gauss\textendash{}Bonnet gravity,
Eur. Phys. J. C \textbf{82}  (2022) 180, arXiv:2108.07454. 

\bibitem{Gammon:2023uss}
M.~Gammon, S.~Rourke and R.~B.~Mann,
Quark stars with a unified interacting equation of state in regularized 4D Einstein-Gauss-Bonnet gravity,
Phys. Rev. D \textbf{109} (2024) 024026, arXiv:2309.00703.

\bibitem{Pretel:2021gqq}
J.~M.~Z.~Pretel, A.~Banerjee and A.~Pradhan,
Five-dimensional compact stars in Einstein-Gauss-Bonnet gravity,
Nucl. Phys. B \textbf{1008}, 116702 (2024), arXiv:2107.03859. 

\bibitem{Tangphati:2024war}
T.~Tangphati, I.~Sakalli, A.~Banerjee and A.~Pradhan,
Anisotropic quark stars in $f(R,L_m,T)$ gravity, arXiv:2404.01970.

\bibitem{Haghani:2023nrm}
Z.~Haghani and T.~Harko,
Compact stellar structures in Weyl geometric gravity,
Phys. Rev. D \textbf{107} (2023) 064068, arXiv:2303.10339. 

\bibitem{Olmo:2019flu}
G.~J.~Olmo, D.~Rubiera-Garcia and A.~Wojnar,
Stellar structure models in modified theories of gravity: Lessons and challenges,
Phys. Rept. \textbf{876} (2020) 1-75, arXiv:1912.05202. 

\bibitem{LIGO}
R.~Abbott \textit{et al.} [LIGO Scientific and Virgo],
GW190814: Gravitational Waves from the Coalescence of a 23 Solar Mass Black Hole with a 2.6 Solar Mass Compact Object,
Astrophys. J. Lett. \textbf{896} (2020) 2, L44, arXiv:2006.12611. 

\bibitem{Nathanail:2021tay}
A.~Nathanail, E.~R.~Most and L.~Rezzolla,
GW170817 and GW190814: tension on the maximum mass,
Astrophys. J. Lett. \textbf{908} (2021) 2, L28, arXiv:2101.01735.

\bibitem{Zhang:2020zsc}
N.~B.~Zhang and B.~A.~Li,
GW190814's Secondary Component with Mass 2.50\textendash{}2.67 M $_{\odot}$ as a Superfast Pulsar,
Astrophys. J. \textbf{902} (2020) 1, 38, arXiv:2007.02513. 

\bibitem{Most:2020bba}
E.~R.~Most, L.~J.~Papenfort, L.~R.~Weih and L.~Rezzolla,
A lower bound on the maximum mass if the secondary in GW190814 was once a rapidly spinning neutron star,
Mon. Not. Roy. Astron. Soc. \textbf{499} (2020) 1, L82-L86, arXiv:2006.14601.

\bibitem{Biswas:2020xna}
B.~Biswas, R.~Nandi, P.~Char, S.~Bose and N.~Stergioulas,
GW190814: on the properties of the secondary component of the binary,
Mon. Not. Roy. Astron. Soc. \textbf{505} (2021) 2, 1600-1606, arXiv:2010.02090.

\bibitem{Zhang:2020dfi}
K.~Zhang and F.~L.~Lin,
Constraint on hybrid stars with gravitational wave events,
Universe \textbf{6} (2020) 12, 231, arXiv:2011.05104.

\bibitem{Prasetyo:2021kfx}
I.~Prasetyo, H.~Maulana, H.~S.~Ramadhan and A.~Sulaksono,
2.6\,\,M\ensuremath{\odot} compact object and neutron stars in Eddington-inspired Born-Infeld theory of gravity,
Phys. Rev. D \textbf{104} (2021) 084029, arXiv:2109.05718.

\bibitem{Ji:2024aeg}
P.~Ji, Z.~Li, L.~Yang, R.~Xu, Z.~Hu and L.~Shao,
Neutron stars in the bumblebee theory of gravity,
Phys. Rev. D \textbf{110} (2024) 104057, arXiv:2409.04805.

\bibitem{Liu}
Liu Y, Sang A-F, Yang W, et al. 
Mass-radius and I-Q relationships of neutron stars in Bumblebee gravity (in Chinese). Sci. Sin-Phys. Mech. Astron.,
 \textbf{54}  (2024) 290411.
 
\bibitem{Panotopoulos:2024jtn}
G. Panotopoulos and A. \"Ovg\"un,
Strange Quark Stars and Condensate Dark Stars in Bumblebee Gravity, arXiv:2409.05801. 

\bibitem{Ince}
E. L. Ince, \textit{Ordinary Differential Equations} (Dover Publications, New York, 1956).

\bibitem{Buchdahl:1959zz}
H.~A.~Buchdahl,
General Relativistic Fluid Spheres,
Phys. Rev. \textbf{116} (1959) 1027.

\bibitem{DeTar:1983rw}
C.~E.~DeTar and J.~F.~Donoghue,
BAG MODELS OF HADRONS,
Ann. Rev. Nucl. Part. Sci. \textbf{33} (1983), 235-264.


\bibitem{Gondek-Rosinska:2008zmv}
D.~Gondek-Rosinska and F.~Limousin,
The final phase of inspiral of strange quark star binaries,  arXiv:0801.4829. 

\bibitem{Arbanil:2016wud}
J.~D.~V.~Arba\~nil and M.~Malheiro,
Radial stability of anisotropic strange quark stars,
JCAP \textbf{11} (2016) 012, arXiv:1607.03984. 

\bibitem{Antoniadis:2013pzd}
J.~Antoniadis, P.~C.~C.~Freire, N.~Wex, T.~M.~Tauris, R.~S.~Lynch, M.~H.~van Kerkwijk, M.~Kramer, C.~Bassa, V.~S.~Dhillon and T.~Driebe, \textit{et al.} 
A Massive Pulsar in a Compact Relativistic Binary,
Science \textbf{340} (2013) 6131, arXiv:1304.6875. 

\bibitem{Fonseca:2021wxt}
E.~Fonseca, H.~T.~Cromartie, T.~T.~Pennucci, P.~S.~Ray, A.~Y.~Kirichenko, S.~M.~Ransom, P.~B.~Demorest, I.~H.~Stairs, Z.~Arzoumanian and L.~Guillemot, \textit{et al.}
Refined Mass and Geometric Measurements of the High-mass PSR J0740+6620,
Astrophys. J. Lett. \textbf{915} (2021) 1, L12, arXiv:2104.00880.

\bibitem{Romani:2022jhd}
R.~W.~Romani, D.~Kandel, A.~V.~Filippenko, T.~G.~Brink and W.~Zheng, 
PSR J0952\ensuremath{-}0607: The Fastest and Heaviest Known Galactic Neutron Star,
Astrophys. J. Lett. \textbf{934} (2022) no.2, L17, arXiv:2207.05124. 

\bibitem{Zeldovich}
Y. B. Zeldovich, and I. D. Novikov, Relativistic Astrophysics,
Vol. I: Stars and Relativity, University of Chicago Press, Chicago, 1971.

\bibitem{Chandrasekhar:1964zz}
S.~Chandrasekhar,
The Dynamical Instability of Gaseous Masses Approaching the Schwarzschild Limit in General Relativity,
Astrophys. J. \textbf{140}, 417-433 (1964).



\end{thebibliography}
\end{document}